\documentclass[lettersize,journal]{IEEEtran}
\usepackage{listings}
\usepackage{graphicx}
\usepackage{shortcuts}
\usepackage{xspace}
\usepackage{code}
\usepackage{cite}
\usepackage{subcaption}
\usepackage{footnote}
\usepackage{mathtools}
\usepackage{url}
\usepackage{chngcntr}
\usepackage{listings, color, caption}
\usepackage{threeparttable}
\usepackage{setspace}
\usepackage{multirow}
\usepackage{multicol}
\usepackage{pifont}
\usepackage{soul}
\usepackage{boxedminipage}
\usepackage{booktabs}
\usepackage{multirow}
\usepackage{hyperref}

\usepackage{algorithm,algorithmicx}
\usepackage[noend]{algpseudocode}
\usepackage{indentfirst}
\usepackage{comment}
\graphicspath{ {images/} }
\usepackage{amsfonts}
\usepackage{mathrsfs}
\usepackage{amssymb}
\usepackage{amsthm}
\usepackage{xcolor}
\definecolor{jsonkey}{RGB}{0,128,128}
\usepackage{multicol}

\newcommand{\RN}[1]{%
  \textup{\uppercase\expandafter{\romannumeral#1}}%
}

\algrenewcommand\algorithmicindent{0.7em}%

\title{DefectHunter: A Novel LLM-Driven Boosted-Conformer-based Code Vulnerability Detection Mechanism}

\author{
 \IEEEauthorblockN{Jin Wang\IEEEauthorrefmark{3}\IEEEauthorrefmark{2},  Zishan Huang\IEEEauthorrefmark{3}\IEEEauthorrefmark{2}, Hengli Liu\IEEEauthorrefmark{3}, Nianyi Yang\IEEEauthorrefmark{3}, Yinhao Xiao \IEEEauthorrefmark{3}\IEEEauthorrefmark{1}}

 \IEEEauthorblockA{\IEEEauthorrefmark{3}School of Information Science, Guangdong University of Finance and Economics, Guangzhou, China.}

 \IEEEauthorblockA{\IEEEauthorrefmark{2}These authors contributed equally to this work}

 \IEEEauthorblockA{\IEEEauthorrefmark{1}Corresponding Author
 	\\ Email: 20191081@gdufe.edu.cn}
 
}

\begin{document}
\maketitle
\begin{abstract}
One of the most pressing threats to computing systems is software vulnerabilities, which can compromise both hardware and software components. Existing methods for vulnerability detection remain suboptimal. Traditional techniques are both time-consuming and labor-intensive, while machine-learning-based approaches often underperform when applied to complex datasets, due to their inability to capture high-dimensional relationships. Previous deep-learning strategies also fall short in capturing sufficient feature information. Although self-attention mechanisms can process information over long distances, they fail to capture structural information. In this paper, we introduce DefectHunter, an innovative model for vulnerability identification that employs the Conformer mechanism. This mechanism fuses self-attention with convolutional networks to capture both local, position-wise features and global, content-based interactions. Furthermore, we optimize the self-attention mechanisms to mitigate the issue of excessive attention heads introducing extraneous noise by adjusting the denominator. We evaluated DefectHunter against ten baseline methods using six industrial and two highly complex datasets. On the QEMU dataset, DefectHunter exhibited a 20.62\% improvement in accuracy over Pongo-70B, and for the CWE-754 dataset, its accuracy was 14.64\% higher. To investigate how DefectHunter comprehends vulnerabilities, we conducted a case study, which revealed that our model effectively understands the mechanisms underlying vulnerabilities.
\end{abstract}

\section{Introduction}
\label{sec:introduction}

The escalating reliance on computer systems and the Internet has engendered transformative shifts across various dimensions of human activity. Simultaneously, it has exacerbated vulnerabilities associated with computational infrastructure. According to the U.S. Department of Commerce's National Institute of Standards and Technology (NIST)\footnote{https://www.nist.gov/}, the National Vulnerability Database (NVD) cataloged an unprecedented 18,378 vulnerabilities in 2021, underscoring a concerning landscape in computer security.

Over the years, scholarly research and professional initiatives in cybersecurity have yielded significant advancements in vulnerability identification through various methodologies and tools. Conventional vulnerability identification techniques often employ static and dynamic analyses. In static analysis, experts scrutinize source code, frequently utilizing automated tools like Flawfinder~\cite{Flawfinder} and Findbugs~\cite{Findbugs}. Dynamic analysis involves monitoring a running program to identify potential security issues. Despite their utility, manual methods are labor-intensive and susceptible to human error. Additionally, automated tools may fall short in detecting intricate vulnerabilities that necessitate an in-depth understanding of system architecture and operational logic. With the advent of machine learning technologies, novel approaches like VCCFinder~\cite{VCCFinder-10.1145/2810103.2813604}, which utilizes Support Vector Machines, and Ghaffarian \textit{et al.} conducted a comparison commonly used machine-learning techniques and found that
the Random Forest algorithm outperformed the others.~\cite{Survey-10.1145/3092566} These strategies, however, may lack efficacy when applied to complex, industrial-scale data sets due to insufficient feature capture. The rise of deep learning techniques offers new avenues for vulnerability identification, with graph-based methods like FUNDED~\cite{FUNDED} and Devign~\cite{zhou2019devign}, and semantic-based approaches like Codebert~\cite{CodeBert} and CodeT5~\cite{wang2021codet5}. Moreover, large language models such as GPT-4~\cite{GPT-4} have substantially augmented the potential of semantic-based techniques. Nonetheless, these methods still encounter challenges in effectively fusing structural and semantic data, limiting their industrial applicability.

To address these shortcomings, we introduce DefectHunter, a comprehensive framework comprising three key modules: structural information processing, a pre-trained machine learning model, and the Conformer mechanism~\cite{gulati2020conformer}. The Conformer mechanism synergizes self-attention with convolution, enabling the capture of both localized and extended dependencies in input sequences. This architectural choice facilitates intricate relationship modeling and optimizes computational efficiency, surpassing prior models like the Transformer. DefectHunter processes structural information from code snippets and tokenizes them into feature arrays encapsulating contextual semantics.  Moreover, we have refined the self-attention mechanisms within the Conformer architecture to mitigate the issue of noise introduced by the utilization of softmax in attention mechanisms.~\cite{Atten6447854:online}. DefectHunter has been fully implemented and its source code is publicly available on GitHub.

In our experiment, we employed an array of industrial datasets, as well as more intricate and challenging datasets like FFmpeg and QEMU, to evaluate the performance of our novel vulnerability identification model, DefectHunter. Our model was benchmarked against 10 established baseline methods. The results demonstrate that the incorporation of structural information substantially enhances capacity of DefectHunter for identifying vulnerabilities. Compared to other transformer-based models, DefectHunter excels in terms of ACC. Specifically, for CWE datasets, DefectHunter achieved an ACC that is 14.64\% higher than that of Pongo-70B. In QEMU datasets, the model surpassed Pongo-70B by achieving a 20.62\% improvement in ACC. This research makes the following contributions:
\begin{itemize}
	\item We introduce DefectHunter, a novel vulnerability identification model, fully implemented and publicly available on GitHub~\footnote{https://github.com/WJ-8/DefectHunter}.
	
	\item We utilize Conformer blocks, an innovative architecture that seamlessly combines convolutional layers with self-attention mechanisms, to enhance performance in sequence modeling tasks. The Conformer is adept at efficiently capturing both local contextual information and intricate feature patterns, whereas the self-attention module facilitates long-range interactions.
	
	\item We refined the self-attention mechanisms within the Conformer block to mitigate the problem of excessive attention heads introducing unwarranted noise.
	
	\item We leveraged a pre-trained large language model to extract semantic information from code snippets, leading to a marked improvement in training efficiency.
	
\end{itemize}

\textbf{Paper Organization.} 
The rest of the paper is organized as follows. Section ~\ref{sec:background} presents recently advanced background knowledge of our approach. Section ~\ref{sec:approach} details the design and technical components of the DefectHunter. Section ~\ref{sec:implementation} demonstrates our implementation of DefectHunter. Section ~\ref{sec:eval} reports our evaluation results and case studies on DefectHunter. Section ~\ref{sec:related} outlines the most related work. Section ~\ref{sec:conclusion} concludes the paper with a future research discussion. 
\section{Background}
\label{sec:background} 

\subsection{Graph} We employ both data flow graphs and control flow graphs as inputs to our computational model, thereby offering essential structural insights. These graph-based frameworks constitute the cornerstone for the systematic interpretation of injected code snippets. Graphs confer multiple benefits in the realm of code analysis. Firstly, they encapsulate the inherent structural aspects of code, thereby facilitating a comprehensive understanding of both data dependencies and control flow constructs. This exhaustive viewpoint allows for the discernment of complex relationships within the software ecosystem. Secondly, graphs afford abstraction by distilling the high-level facets of the code and disregarding non-essential elements. This abstraction considerably mitigates the analytical complexity. Lastly, graph-based formulations are naturally compatible with machine learning paradigms and other computational methodologies, thus offering a structured input conducive to predictive modeling.

\subsection{Self-Attention Based Methods}
Self-attention based methods offer various trade-offs between computational efficiency and the ability to capture long-range dependencies, making them valuable tools in the fields of deep learning and natural language processing.

\textbf{Transformer:} Introduced by Vaswani \textit{et al.}~\cite{DBLP:journals/corr/VaswaniSPUJGKP17}, the Transformer architecture surmounts the limitations intrinsic to both recurrent and convolutional neural networks in capturing long-range dependencies. This makes it particularly apt for tasks that require the interpretation of sequential or structured data, such as machine translation and natural language understanding.

Central to the Transformer is the self-attention mechanism, which endows the model with the capability to selectively focus on various portions of the input sequence during predictive tasks. This mechanism allows the model to effectively grasp both local and global contexts, thereby ensuring high parallelizability and reducing training durations. The architecture is constituted by an encoder-decoder design, featuring multiple layers of both self-attention and feedforward neural networks.

\textbf{Conformer:} Developed by Anmol Gulati \textit{et al.}~\cite{gulati2020conformer}, the Conformer architecture enhances the Transformer by ameliorating some of its deficiencies. While the Transformer excels in the capture of global contexts, it is challenged by long sequences due to its quadratic computational complexity. The Conformer rectifies this by ingeniously combining convolutional and self-attention mechanisms. A noteworthy innovation within the Conformer is the convolutional feed-forward module, facilitating efficient capture of local dependencies. This architectural nuance allows the Conformer to sustain high performance levels even in scenarios involving extended sequences. Moreover, the Conformer retains the prowess of the self-attention in effectively capturing global contexts.

\subsection{Large Lanuage Models} Over the past decade, large language models have emerged as revolutionary entities in the field of natural language processing. These models are characterized by their gargantuan parameter counts, often ranging from hundreds of millions to billions, enabling them to apprehend complex linguistic structures and produce contextually pertinent text. Prominent examples include the GPT series by OpenAI, BERT by Google, and T5 by Microsoft.

These models hold significant potential in the domain of code analysis. When supplied with code snippets, they are capable of aiding developers in a variety of tasks, such as code auto-completion, bug identification, and refactoring recommendations. Their intrinsic ability to understand both the context and syntax of programming languages renders them invaluable assets in software development endeavors.
\section{Design of DefectHunter}
\label{sec:approach}
This section describes the architecture of DefectHunter, which comprises three fundamental components: structural information processing, a pre-trained model, and the Conformer mechanism. The workflow begins with the extraction of structural information using an open-source tool. A semantic information feature matrix is subsequently generated through a specialized large language model designed for code embedding. The Conformer mechanism is then applied to distill vulnerability features from both the structural and semantic data. Adjustments have been made to the self-attention mechanisms within the Conformer to address the issue of superfluous attention heads contributing to extraneous noise. Ultimately, a multi-layer perceptron is employed to ascertain the presence or absence of vulnerabilities.

\begin{figure*}[htbp]
	\includegraphics[width=190mm]{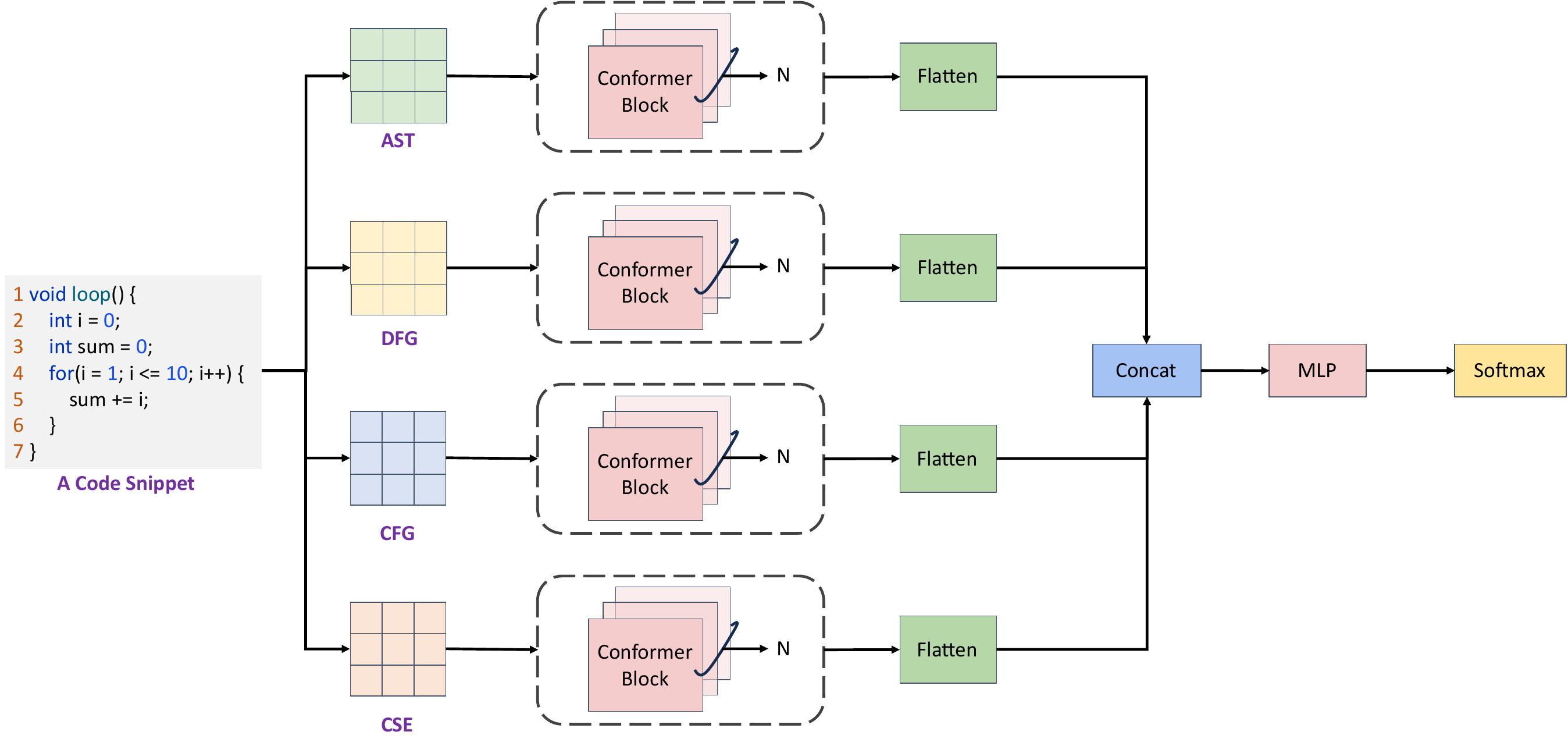}
	\centering
	\caption{Design of the DefectHunter Model}
	\label{fig:model}
\end{figure*}

\subsection{Structural Information}

Structural information serves as a foundational component in the architecture of DefectHunter, generating a diverse set of graphs such as Abstract Syntax Trees (AST), Control Flow Graphs (CFG), and Data Flow Graphs (DFG) from source code.

The AST provides a hierarchical framework that delineates the abstract syntax of a program, with each node symbolizing a syntactic construct and edges representing hierarchical relationships. This model substantially aids in the analysis and understanding of code. The CFG outlines possible execution pathways within a program, employing nodes to signify program constructs and edges to mark transitions contingent on branching operations. This graph clearly identifies entry and exit points, furnishing a visual guide to the sequence of program execution. Similarly, the DFG captures the interplay of data and dependencies among operations, spotlighting variable instantiation, modification, and usage. In the DFG, nodes stand for variables or operations, while edges signify data dependencies. Collectively, AST, CFG, and DFG contribute to a comprehensive understanding of a program structure, logic, and data flow, thereby enhancing the capabilities of DefectHunter in defect identification and analysis.

\subsection{Code Sequence Embedding (CSE)}

At the heart of Code Sequence Embedding (CSE) lies the use of pre-trained models to generate word embeddings. Each token is transformed into a feature vector, referred to as a contextual token representation, through a pre-trained model. Distinct from traditional word-embedding methods, which necessitate substantial training, CSE leverages pre-training techniques to mitigate the risk of overfitting in situations with limited or biased training data. Consequently, pre-training affords more relevant features.  $x_i$ denotes a given piece of code, and the acquired representation $P_i$ through Eq.\ref{eq:CSE} is as follows:
\begin{equation}
	\label{eq:CSE}
	M_i = \text{model}(x_i)
\end{equation}
$\text{model}$ represents the mathematical expression of the employed pre-trained model.

\subsection{Conformer}

The Conformer employs sinusoidal positional embeddings to encode the input sequences, thereby supplying the model with essential information about the ordering and relative positions of sequence elements. The sinusoidal encodings are generated according to the formulas below:

\begin{align}
	\text{for even } i \notag \\
	\text{pos}_i &= \left[\sin\left(\frac{\text{pos}}{10000^{2i/d}}\right), \cos\left(\frac{\text{pos}}{10000^{2i/d}}\right)\right] \\
	\text{for odd } i \notag \\
	\text{pos}_i &= \left[\cos\left(\frac{\text{pos}}{10000^{2i/d}}\right), \sin\left(\frac{\text{pos}}{10000^{2i/d}}\right)\right] 
\end{align}
where $i$ represents the position within the sequence, $d$ is the dimensionality of the sinusoidal encoding, and $pos_i$ corresponds to the sinusoidal encoding of the $i-th$ token. Prior to the self-attention process, the input sequence is element-wise multiplied by these sinusoidal encodings.

\begin{figure}[htbp]
	\includegraphics[width=80mm]{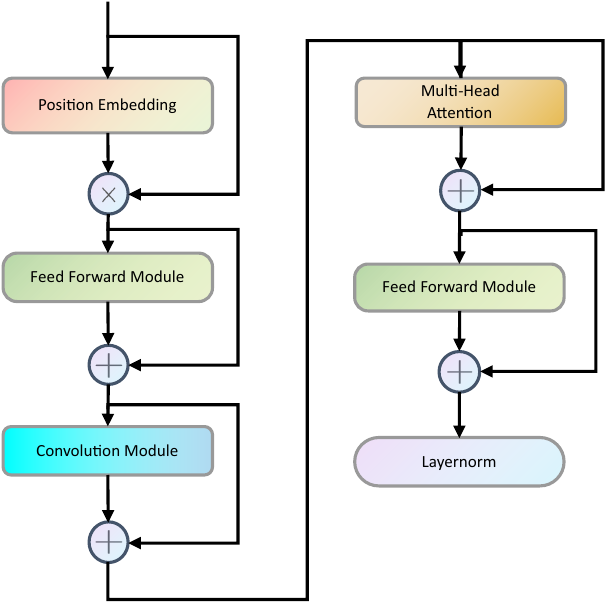}
	\centering
	\caption{Conformer Block}
	\label{fig:block}
\end{figure}

The Conformer is a sophisticated architectural design that synergizes the capabilities of Convolutional Neural Network (CNN) and self-attention mechanisms to augment sequence modeling tasks. This amalgamation enables the Conformer to efficiently capture both local and global dependencies in a sequence, thereby overcoming some limitations inherent in conventional Transformer models. Compared to traditional Transformers, the Conformer is adept at identifying both position-wise local features and content-based global interactions. The incorporation of a CNN module allows for effective extraction of local context and intricate feature patterns, while the self-attention mechanism is responsible for capturing long-range dependencies.

The Conformer block comprises three primary modules: a convolutional module, a self-attention module, and two feed-forward neural networks, as illustrated in Fig.~\ref{fig:block}. Each module performs a vital function in processing various aspects of the input sequence, collectively contributing to the Conformer efficacy. We have also implemented specific alterations to the traditional Conformer block. Rather than solely utilizing sinusoidal positional encodings for multi-head attention, we combine these encodings with the input matrix before feeding the result into a fully connected layer. This adjustment optimizes the encoding process at the input stage of the Conformer model.

\subsubsection{Convolutional Module}

The Convolutional Module within the Conformer architecture employs CNN to effectively capture local dependencies in sequential data. This module is comprised of an array of convolutional layers, which operate concurrently and hierarchically to distill pertinent features from the input sequence. Mathematically, the output of a given convolutional layer can be formulated as:
\begin{equation}
	Conv(x) = ReLU(BatchNorm(W * x + b))
\end{equation}

where $Conv(\cdot)$ denotes the convolutional layer, $ReLU(\cdot)$ represents the rectified linear unit activation function, $BatchNorm(\cdot)$ denotes batch normalization, $W$ represents the convolutional kernel, $x$ is the input sequence, and $b$ represents the bias term.

\subsubsection{Self-Attention Module}

The self-attention module employs multi-head self-attention to capture global interdependencies within the input sequence. This is achieved by simultaneously attending to different portions of the sequence. Specifically, attention scores are computed using the dot product of the query ($Q$) and key ($K$) vectors, normalized by the square root of the dimension of the key/query space ($d_k$). These scores are subsequently passed through a softmax function to derive weights representing the importance of each sequence element relative to others. These weights are utilized to modulate the value ($V$) vectors, culminating in the output of the module. Mathematically, the multi-head self-attention operation is expressed as:
\begin{equation}
	\!\!\!\!\text{MHSA}(Q, K, V) = \text{Concat}(\text{head}_1, \text{head}_2, \cdots, \text{head}_h)W_O
\end{equation}
where the $h$ heads represent different sets of learned projection weight matrices $W_Q$, $W_K$, and $W_V$ for the queries, keys, and values respectively. Each head operates on transformed versions of the input sequences, as follows:

\begin{align}
	\text{head}_i = \text{Attention}(QW_{Qi}, KW_{Ki}, VW_{Vi}) \\
	\label{eq:attention}
	\text{Attention}(Q_i, K_i, V_i) = \text{Softmax}\left(\frac{Q_iK_i^T}{\sqrt{d_k}}\right) V_i
\end{align}
\subsubsection{Attention Calculation Modification}
A critical issue arises in Eq.\ref{eq:attention}. The operation $QK^T$ aims to establish correlations among token (embedding) vectors positioned differently, essentially constructing a square correlation matrix. This matrix is composed of dot-product values, scaled by $\frac{1}{\sqrt{d}}$, wherein each column and row corresponds to a specific token position. Subsequently, each row of this square matrix undergoes a softmax operation, generating probabilities that serve as a mechanism for combining the value vectors within the matrix $V$. The resultant probability-weighted $V$ matrix is then added to the initial input vector. This cumulative sum is propagated through the neural network for further layers of processing.

In the context of multi-head attention, this entire procedure is repeated multiple times in parallel for each layer. In essence, the embedding vector is partitioned, and each attention head employs the complete vector's information to annotate a distinct and non-overlapping section of the resulting vector.

However, the use of softmax introduces a drawback, compelling each attention head to annotate even when it lacks relevant information. While effective for discrete selection tasks, softmax is less suited for optional annotation, especially when the output is summative. This challenge is exacerbated in multi-head attention, where specialized heads are less likely to contribute compared to their general-purpose counterparts. This results in unnecessary noise, undermining the  performance of the model. 

To rectify this issue, inspired by Evan Miller~\cite{Atten6447854:online}, we modify the denominator in the softmax equation by adding 1. This ensures a positive derivative and a bounded output, thereby stabilizing the model.
\begin{align}
	\label{eq:fix-attention}
	\text{Attention}(Q_i, K_i, V_i) = \text{Softmax}\left(\frac{Q_iK_i^T}{1+\sqrt{d_k}}\right) V_i
\end{align}

\subsubsection{Feed-Forward Neural Network}

The feed-forward neural network in Conformer introduces non-linear transformations to capture complex relationships between features. It consists of two linear layers, with a ReLU activation function in between, as defined by the following equation:
\begin{equation}
	\text{FFN}(x) = \text{ReLU}(W_2(\text{ReLU}(W_1x + b_1)) + b_2)
\end{equation}
where $\text{FFN}(\cdot)$ represents the feed-forward neural network, $W_1$, $W_2$, $b_1$, and $b_2$ are weight and bias terms, respectively.

\section{Implementation}
\label{sec:implementation}
This section provides a comprehensive description of the DefectHunter implementation.

\subsection{Building Graph}
We utilize the tree-sitter library\footnote{https://tree-sitter.github.io/tree-sitter/} to generate AST from source code snippets. Our methodology consists of parsing the source code using the C language parser supplied by tree-sitter, which results in AST objects. These ASTs are subsequently converted into Graphviz Dot format. From these Dot representations, edges are extracted and transformed into adjacency matrices. For CFG and DFG, the preliminary phase involves parsing the code using tree-sitter. AST nodes are traversed according to specific rules to generate either a CFG, which contains nodes corresponding to program execution statements, or a DFG, which includes nodes that represent variable declarations and modifications.
\subsection{Building CSE}
\begin{figure}[htbp]
	\includegraphics[width=45mm]{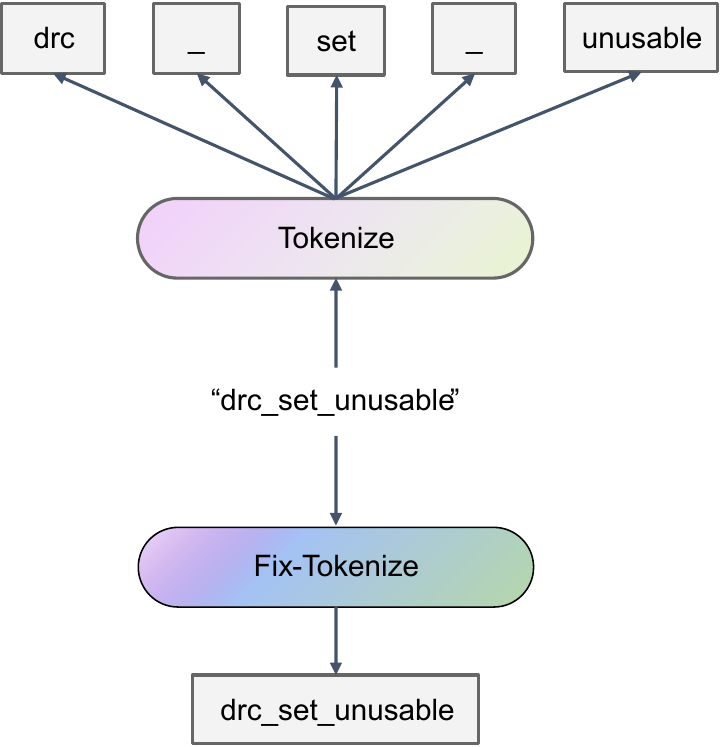}
	\centering
	\caption{Tokenizing Result}
	\label{fig:tokenizer}
\end{figure}

We leverage the transformer library~\cite{wolf-etal-2020-transformers} to import UniXcoder~\cite{guo2022unixcoder}, a unified cross-modal pre-training model for programming languages. In our experiments, we encountered challenges in tokenizing the source code, which led to imprecise word embeddings. As illustrated in Fig.\ref{fig:tokenizer}, using a function name from the QEMU dataset—specifically, \texttt{drc\_set\_unusable}—we observed that UniXcoder tokenizes it inaccurately. To mitigate this issue, we utilized the Natural Language Toolkit (NLTK)~\cite{bird2009natural} to assist in the tokenization process. Furthermore, we generated a specialized word list that UniXcoder uses during tokenization, ensuring the preservation of specific words. The source code is then fed into UniXcoder to generate Code Source Embeddings (CSEs) as matrices. It is important to note that UniXcoder supports up to 768 input tokens, necessitating the configuration of the \texttt{max\_length} parameter in the tokenizer.

\subsection{Network Implementation}
We implied these networks through custom layers of Keras~\cite{chollet2015keras}.
\subsubsection{Multi-Head Self-Attention}
We introduce a multi-head self-attention layer that accepts the embedding dimension, the number of heads, and an optional dropout rate as its parameters. This layer comprises multiple components: \texttt{query\_dense}, \texttt{key\_dense}, and \texttt{value\_dense} perform linear transformations on the input, while the attention mechanism computes the attention scores. The \texttt{separate\_heads} function reshapes and transposes the input to generate multiple heads, and the \texttt{call} function orchestrates these components to yield the final output. The attention mechanism follows the formula detailed in Eq.\ref{eq:fix-attention}, where the \texttt{query}, \texttt{key}, and \texttt{value} tensors are multiplied to calculate the score. This score is scaled by $1+\sqrt{d_k}$ ,and the softmax function is subsequently applied to produce the attention weights. Finally, the attention output is obtained by multiplying the weights with the value tensor.
\subsubsection{Sinusoidal Position Encoding}
To integrate sinusoidal position encoding into DefectHunter, we implement a separate layer named Sinusoidal Position Embedding. This layer processes the input sequence by applying sinusoidal position embeddings, which are computed based on generated position IDs and indices. The input sequence is then multiplied by these embeddings.

\section{Evaluation}
\label{sec:eval}

\subsection{Experimental Setup}
In this section, we first give a presentation of the preparatory measures taken before the experiments were carried out, including the setup of the environment and the selection of the model training parameters. Then, we provide an elaborative description of the datasets employed, notably focusing on the two datasets we curated. One is tailored for pre-training a large language model (LLM), and the other is specifically tailored for training the DefectHunter model.

We proceed to present comparative experiments against established benchmarks, thereby demonstrating the efficacy of DefectHunter in defect detection. Additionally, we conduct a decomposition analysis to investigate the relationship between model performance and key influencing factors. In concluding our evaluation, we provide an in-depth case study exploring the proficiency of DefectHunter in identifying software vulnerabilities and its potential real-world applications.
\subsubsection{Performance Metrics}

We examined the performance of the model in terms of F1-Score and Accuracy, and here is a brief description of these performance metrics.

\textbf{Accuracy}: This metric measures the proportion of correctly classified samples to the total number of samples. Mathematically, it's calculated as:

\[ \text{Accuracy} = \frac{\text{Number of Correct Predictions}}{\text{Total Number of Predictions}} \]

\textbf{F1-Score}: F1-Score is the harmonic mean of precision and recall, offering a balance between the two metrics. It is particularly useful when the classes are imbalanced. Mathematically, it is the harmonic mean of precision and recall:

\[ F1 = 2 \times \frac{\text{Precision} \times \text{Recall}}{\text{Precision} + \text{Recall}} \]

\textbf{Precision}: Precision quantifies the proportion of positive identifications that were actually correct. It focuses on the rate of false positives. Mathematically:

\[ \text{Precision} = \frac{\text{True Positives}}{\text{True Positives} + \text{False Positives}} \]

\textbf{Recall}: Recall quantifies the proportion of actual positive cases that were correctly identified. It focuses on the rate of false negatives. Mathematically:

\[ \text{Recall} = \frac{\text{True Positives}}{\text{True Positives} + \text{False Negatives}} \]

\subsubsection{Environment Configuration}
The computational environment for this study is configured with a high-performance workstation, outfitted with dual AMD EPYC 7543 processors, each containing 32 cores. The system is supplemented by 320GB of RAM and an array of four NVIDIA A40 GPUs, each equipped with 48GB of VRAM. On the software front, the implementation utilizes TensorFlow v2.7.0 in conjunction with Keras v2.7.0. The employed neural network architecture incorporates a Conformer encoder composed of 12 Conformer blocks. Each block consists of eight attention heads and a fully connected network with 1,024 dimensions. Optimization is carried out using the Adam optimizer, configured with a learning rate of $1e-5$  and a batch size of 64. Experiments on the FFmpeg and QEMU datasets are conducted over 50 epochs, whereas the training on the CWE dataset is restricted to 30 epochs. The model's architecture comprises a substantial 0.19 billion parameters.
\subsection{Dataset}

\begin{table*}[htbp]
	\centering
	\caption{Dataset Information}
	\label{table:dataset}
	\begin{tabular}{l|llll|ll}
		\hline
		\textbf{Project} & \textbf{Training Set} & \textbf{Validation Set} & \textbf{Test Set} & \textbf{Total} & \textbf{Vul} & \textbf{Non-Vul} \\ \hline
		\textbf{FFmpeg}  & 3958                  & 462                     & 499               & 4919           & 2438         & 2481             \\
		\textbf{QEMU}    & 10903                 & 1378                    & 1318              & 13600          & 5687         & 7913             \\
		\textbf{CWE-362}    & 451                 & 56                   & 57              & 564          & 189         & 375             \\
		\textbf{CWE-476}    & 1298                 & 162                   & 163              & 1623          & 396         & 1227             \\
		\textbf{CWE-754}    & 4034                 & 504                    & 505              & 5043          & 1359         & 3684             \\
		\textbf{CWE-758}    & 1089                 & 136                    & 137              & 1362          & 367         & 995             \\
		\textbf{CWE-basic}    & 222                 & 0                    & 0              & 222          & 222         & 0             \\ \hline
	\end{tabular}
\end{table*}

To validate the effectiveness of our proposed model, we utilize an extensive set of six datasets, each of which has been previously employed in state-of-the-art research~\cite{zhou2019devign, FUNDED, wang2021codet5}. These datasets fall into two primary categories. The first category consists of standard datasets derived from the Software Assurance Reference Dataset (SARD)~\cite{NISTSoft48:online}, specifically CWE-362, CWE-476, CWE-754, and CWE-758. These datasets cover a spectrum of software vulnerabilities, such as race conditions, null pointer dereferences, improper handling of exceptional conditions, and reliance on undefined behavior. To thoroughly evaluate the performance of our model performance, we introduce a second category that includes more complex datasets: FFmpeg and QEMU. FFmpeg serves as a multimedia processing framework, while QEMU operates as an emulator; both play crucial roles in software development and feature complex codebases. Unlike the CWE datasets, vulnerabilities in FFmpeg and QEMU are inherently complex, thereby imposing stringent requirements on our model's vulnerability detection capabilities. Comprehensive details of the datasets, including the number of vulnerabilities, the distribution across training, validation, and testing subsets, and the overall dataset size, are presented in Table \ref{table:dataset}.

\subsubsection{Dataset For LLM}
\begin{figure*}[htbp]
	\includegraphics[width=150mm]{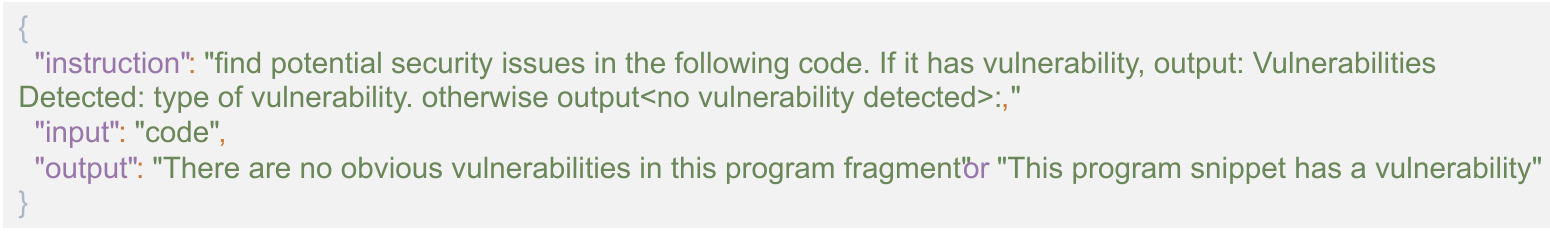}
	\centering
	\caption{Alpaca Dataset Format}
	\label{fig:format}
\end{figure*}

We have collected a dataset, termed \texttt{CWE-basic}, sourced from the Common Weakness Enumeration Specification to augment the capabilities of LLM in identifying a broader array of vulnerabilities. This dataset encompasses 77 distinct types of Common Weakness Enumeration (CWE) vulnerabilities. Specifically, we employed the CWE instances cataloged by the National Vulnerability Database and constructed our dataset based on the examples provided therein. To automate this process, we developed a web crawler\footnote{https://github.com/Yuzu815/nist-vul-crawler}. Each example in the dataset was segmented into four components: a code snippet, a vulnerability indicator, the associated programming language, and a contextual description. Subsequently, we isolated code snippets featuring vulnerabilities and formatted them for LLM training as depicted in Fig.\ref{fig:format}.
Specifically, the LLM requires a specialized input format, denoted as the \texttt{Alpaca} format. To adhere to this requirement, the FFmpeg, QEMU, and CWE datasets must be modified to align with the format illustrated in Figure \ref{fig:format}.


\subsection{Baseline Methods}

We evaluate DefectHunter by contrasting its performance with ten disparate baseline methods, encompassing traditional neural network architectures such as CNN and GNN, as well as Transformers and LLMs.

\subsubsection{CNN~\cite{kim-2014-convolutional}}
We utilize Word2Vec for code embedding, subsequently applying a CNN for feature extraction. The model culminates with a singular dense classification layer to achieve categorization.

\subsubsection{CodeBERT~\cite{CodeBert}}
CodeBERT is a bimodal pre-trained model optimized for both programming and natural languages. Built upon a Transformer architecture, it is trained with a hybrid objective function incorporating a replaced token detection pre-training task. CodeBERT can generate generalized representations applicable to various natural language and programming language tasks. For vulnerability identification, its output is integrated into a multi-layer fully connected network.
\subsubsection{SELFATT~\cite{DBLP:journals/corr/VaswaniSPUJGKP17}}
We tokenize source code using a tokenizer, yielding a sequence of code tokens. Employing a multi-head self-attention mechanism, we process the sequence, culminating in a fully connected layer for classification.

\subsubsection{Devign~\cite{zhou2019devign}}
Devign is a model based on GNN that aims to identify software vulnerabilities within source code. The model employs graph-level classification techniques facilitated by a diverse set of code semantic representations. A unique aspect of Devign is the inclusion of a novel Convolutional module, which adeptly extracts salient features from pre-learned node representations to augment the graph-level classification. This methodology is predicated on the use of comprehensive and multifaceted code semantic representations. Utilizing a GNN-based framework enables the discernment of intricate relationships and patterns in code structures, often serving as indicators of vulnerabilities.

\subsubsection{VulDeepecker~\cite{li2018vuldeepecker}}
VulDeePecker is a deep learning-based framework for automated software vulnerability detection. It utilizes code gadgets—clusters of semantically related lines of code—converted into vectors as its representation methodology.

\subsubsection{FUNDED~\cite{FUNDED}}
FUNDED is a novel learning framework for building effective vulnerability detection models. Leveraging graph neural networks,  FUNDED employs a unique graph-based learning approach to capture program control, data, and call dependencies. Unlike previous methods, which often consider programs sequentially or as untyped graphs, FUNDED utilizes a graph representation of source code. This representation connects statements through relational edges, encapsulating syntax, semantics, and flows for improved code representation in vulnerability detection.

\subsubsection{DeepVulSeeker~\cite{WANG202315}}
DeepVulSeeker is a framework for vulnerability identification that integrates code graph structures and semantic features. It overcomes the limitations of existing methods by utilizing Graph Representation Self-Attention and pre-training mechanisms to achieve high accuracy.
\subsubsection{Codex~\cite{chen2021evaluating}}
Codex is an advanced GPT-based language model fine-tuned on publicly accessible GitHub code. Although proficient in Python programming, Codex has limitations in understanding complex docstrings and in mapping operations to variables effectively.
\subsubsection{Pongo-13B}
Pongo-13B leverages the llama2-13b~\cite{touvron2023llama} base model, fine-tuned through 600 iterations using the QLORA~\cite{dettmers2023qlora} technique. Employing four A40 48GB GPUs, it is trained on intricate code segments from the FFmpeg and QEMU datasets, supplemented by the CWE-basic dataset. Pongo-13B is open-sourced on Hugging Face\footnote{https://huggingface.co/wj2003/Pongo-13B}.

Guiding the model responses is a meticulously structured prompt: ``\texttt{Find potential security issues in the following code. If it has a vulnerability, output: Vulnerabilities Detected: type of vulnerability. otherwise output<no vulnerability detected>:}''

Under this prompt, the model outputs exhibit discernible patterns: when vulnerabilities are detected in the provided code, the model generates the response ``\texttt{This program snippet has a vulnerability.}'' followed by the specific type of vulnerability detected. In cases where the code is deemed free of vulnerabilities, the model produces the response ``\texttt{This program snippet has no vulnerability detected.}''

\subsubsection{Pongo-70B}
Built upon the llama2-70b model, Pongo-70B undergoes 800 training steps, maintaining other parameters consistent with Pongo-13B. It is also open-sourced on Hugging Face\footnote{https://huggingface.co/wj2003/Pongo-70B}.

\subsection{Experimental Results}

\begin{table*}[]
	\centering
	\caption{Experimental Results on CWE}
	\label{table:cwe_result}
	\begin{tabular}{cccccccccccc}
		\hline
		\multirow{2}{*}{\textbf{Method}} & \multicolumn{2}{c}{\textbf{CWE-476}} &  & \multicolumn{2}{c}{\textbf{CWE-758}} &  & \multicolumn{2}{c}{\textbf{CWE-362}} & \textbf{} & \multicolumn{2}{c}{\textbf{CWE-754}} \\ \cline{2-3} \cline{5-6} \cline{8-9} \cline{11-12} 
		& ACC               & F1               &  & ACC               & F1               &  & ACC               & F1               &           & ACC               & F1               \\ \hline
		CNN~\cite{kim-2014-convolutional}                 & 0.5819            & 0.3812           &  & 0.6731            & 0.2031           &  & 0.6162            & 0.5083           &           & 0.7371            & 0.343            \\
		CodeBERT~\cite{CodeBert}    &     0.9083    &   0.8387       &  &   0.8356      &  0.7177      &  &   0.7691       &     0.8695     &           &    0.5246       &   0.4706     \\
		SELFATT~\cite{DBLP:journals/corr/VaswaniSPUJGKP17}              & 0.8401            & 0.6213           &  & 0.9121            & 0.8656           &  & 0.6627            & 0.0307           &           & 0.9219            & 0.8769           \\
		Devign~\cite{zhou2019devign}            &    0.8250      &   0.5333      &  &   0.8822      &   0.8164     &  &  0.8571      &   0.8148      &           &  0.9291      &  0.8794         \\
		VulDeepecker~\cite{li2018vuldeepecker}                &0.8070              & 0.8932         &  &  0.7887         &  0.8818      &  &  0.9500      &    0.9048 &           &    0.9574         & 0.9351            \\
		FUNDED~\cite{FUNDED}                   & 0.8889            & 0.9087           &  & 0.9583            & 0.9615           &  & 0.9446            & 0.9521           &           & 0.9286            & 0.9331           \\
		DeepVulSeeker~\cite{WANG202315}              & 0.9080            & 0.8052           &  & 0.9927            & 0.9859           &  & 0.9123            & 0.8387           &           & 0.9999            & 0.9999           \\
		Pongo-13B             & 0.7485            & 0.6239           &  & 0.7591            & 0.3529           &  & 0.5965            & 0.4889           &           & 0.6950            & 0.2735           \\
		Pongo-70B            & 0.8405            & 0.6829           &  & 0.7445            & 0.05405           &  & 0.7368            & 0.5714           &           & 0.8535            & 0.6606           \\
		DefectHunter            & 0.9141            & 0.8293           &  & 0.9980            & 0.9922           &  & 0.9474            & 0.9518           &           & 0.9999            & 0.9999           \\ \hline
	\end{tabular}
\end{table*}

\begin{table}[ht]
	\centering
	\caption{Experimental Results on FFmpeg and QEMU}
	\resizebox{\columnwidth}{!}{
		\label{table:ffmpeg_result}
		\begin{tabular}{@{}cccccc@{}}
			\toprule
			\multirow{2}{*}{\textbf{Method}} & \multicolumn{2}{c}{\textbf{FFmpeg}} &  & \multicolumn{2}{c}{\textbf{QEMU}} \\ \cmidrule(l){2-6} 
			& ACC              & F1               &  & ACC             & F1              \\ \midrule
			CNN~\cite{kim-2014-convolutional}                & 0.5493           & 0.5341           &  & 0.5892          & 0.1330          \\
			VulDeePecker~\cite{li2018vuldeepecker}                &   0.5637         &  0.5702        &  &   0.5980      &   0.6141      \\
			SELFATT~\cite{DBLP:journals/corr/VaswaniSPUJGKP17}                  & 0.5710           & 0.5275           &  & 0.6049          & 0.4904          \\
			CodeBERT~\cite{CodeBert}           &  0.5264           &  0.6138           &  &  0.5354       &   0.7126       \\
			Devign~\cite{zhou2019devign}                & 0.5904           & 0.6015           &  &  0.6039        &  0.3244         \\
			FUNDED~\cite{FUNDED}              & 0.5420           & 0.6800           &  & 0.5970        & 0.7480       \\
			DeepVulSeeker~\cite{WANG202315}                           & 0.6354           & 0.6703           &  & 0.6409          & 0.5293          \\ 
			Pongo-13B                           & 0.5991           & 0.6428           &  & 0.4238          & 0.5945          \\ 
			Pongo-70B                           & 0.6132           & 0.5832           &  & 0.4397          & 0.6003          \\
			Codex~\cite{chen2021evaluating}                           & \multicolumn{5}{c}{F1: 0.5919}          \\
			DefectHunter                           & 0.6653           & 0.7023           &  & 0.6459          & 0.7095          \\ \bottomrule
		\end{tabular}
	}
\end{table}

In this subsection, we provide a detailed analysis of the experimental results obtained from evaluating our proposed model, DefectHunter, on different datasets. The results are presented in Tables \ref{table:ffmpeg_result}, \ref{table:cwe_result}. We compare our model against 10 baseline methods, comprising both traditional neural networks (CNN, GNN) and modern transformer-based models (SELFATT, CodeBERT, and DeepVulSeeker), as well as large language models (LLM) such as Codex, Pongo-13B and Pongo-70B.
\begin{itemize}
	\item The inclusion of structural information significantly enhances the vulnerability identification performance of our model. Comparing the results of CNN, a traditional network, with Devign, we observe substantial improvements in both accuracy and F1 scores across the datasets. This underscores the effectiveness of learning local code characteristics through structural information. Moreover, FUNDED outperforms CodeBERT, indicating that structural information contributes positively to vulnerability identification.
	\item The advantages of attention mechanisms and pre-training are manifest. SELFATT demonstrates superior performance in terms of accuracy on the FFmpeg and QEMU datasets compared to VulDeePecker. Furthermore, CodeBERT, as a pre-trained model, consistently surpasses SELFATT across a multitude of metrics, thus underscoring the benefits of leveraging pre-trained representations.
	
	\item A salient observation is the significant contribution of the Conformer block to transformer-based models. Upon comparing DefectHunter with other transformer-based models such as CodeBERT, SELFATT, and DeepVulSeeker, it becomes evident that DefectHunter consistently outperforms them in terms of accuracy. This superiority can be attributed to the specialized architecture of the Conformer block, which integrates deformable convolutions and self-attention mechanisms, thereby enhancing the capture of intricate structural information in code.
	
	\item Despite the prodigious advancements of LLMs in various natural language processing endeavors, our results indicate that for vulnerability identification, models with smaller parameter sizes like DeepVulSeeker and DefectHunter offer distinct advantages. These smaller models are more resource-efficient, necessitating lower hardware requirements and facilitating easier deployment.
\end{itemize}

\subsection{Ablation Study}

\begin{table*}[]
	\caption{Ablation Study}
	\label{table:ablation}
	\centering
	\begin{tabular}{ccccc}
		\hline
		\multirow{2}{*}{Method} & \multicolumn{2}{c}{FFmpeg} & \multicolumn{2}{c}{Qemu} \\ \cline{2-5} 
		& ACC          & F1          & ACC         & F1         \\ \hline
		DefectHunter               & 0.6653       & 0.7023      & 0.6459      & 0.7095     \\
		DefectHunter w/o AST       & 0.6232       & 0.6713      & 0.6338      & 0.3970     \\
		DefectHunter w/o DFG       & 0.6253       & 0.6492      & 0.6331      & 0.4550     \\
		DefectHunter w/o CFG       & 0.6333       & 0.6376      & 0.6293      & 0.5606     \\
		DefectHunter w/o Conformer       & 0.6052       & 0.6099      & 0.6118      & 0.3663     \\
		DefectHunter w/o Attention-modified       & 0.6112       & 0.6381      & 0.5944      & 0.5078     \\
		DefectHunter w/o LLM       & 0.5230       & 0.5131      & 0.5777      & 0.5831     \\ \hline
	\end{tabular}
\end{table*}
To comprehensively understand the impact of individual modules within our proposed model, we performed an ablation study. The study focused on three critical components: the Conformer module, the Attention-modified layer, and a LLM. We tested the performance of our model on two tasks, FFmpeg and Qemu, using ACC and F1-score as performance metrics. The results are summarized in Table \ref{table:ablation}.

\subsubsection{Conformer Module}
The Conformer module is crucial for enhancing both self-attention and convolution layers for better feature extraction and representation. When we removed this module, the performance for FFmpeg dropped notably from an ACC of 0.6653 to 0.6052 and an F1-score of 0.7023 to 0.6099. The drop was even more substantial for the Qemu task, where ACC reduced from 0.6459 to 0.6118 and the F1-score plummeted from 0.7095 to 0.3663. This steep decline, especially in the F1-score for Qemu, underscores the importance of the Conformer module.

\subsubsection{Attention-modified Layer}
The incorporation of our Attention-modified layer enhances the model capability to prioritize salient features in the input data. When this layer was ablated, there was a decline in accuracy (ACC) from 0.6653 to 0.6112 and a corresponding reduction in F1-score from 0.7023 to 0.6381 in the case of FFmpeg. For Qemu, a similar decline was observed: the ACC fell from 0.6459 to 0.5944, and the F1-score decreased from 0.7095 to 0.5078. Although the deterioration in performance was evident, it was less pronounced compared to the removal of the Conformer module, suggesting that the Attention-modified layer, while valuable, may not be as indispensable as the Conformer module.

\subsubsection{LLM}
The LLM aims to provide a structured understanding of the input data, interpreting it in a way that adds meaningful context for the task at hand. Eliminating the LLM resulted in a severe drop in performance. Specifically, for FFmpeg, the ACC plummeted from 0.6653 to 0.5230, and the F1-score sank from 0.7023 to 0.5131. Similarly, in the Qemu task, ACC reduced from 0.6459 to 0.5777 and F1-score from 0.7095 to 0.5831. These considerable decreases highlight the critical role that the LLM plays in both tasks.

\subsubsection{Structural Information}
To delve into the contribution of different structural components, we performed ablation tests on our model. The results, as summarized in the provided table, clearly demonstrate the pivotal role of structural information in enhancing performance. The removal of any of the graphs, namely AST, CFG or DFG, led to consistent declines in both accuracy and F1-score. This underscores the importance of preserving the structural integrity of the model.

In conclusion, our ablation study vividly shows that each module in the DefectHunter model significantly contributes to its overall performance. The removal of any one component led to a considerable decline in both ACC and F1-scores across tasks, emphasizing the importance of the synergistic relationship among them.

\subsection{Case Study}
We conducted a case study to explore how DefectHunter understands vulnerability.
\subsubsection{Case 1}
\begin{figure*}[!htb]
	\centering
	\subfloat[Vulnerable\label{fig:case1_vul}]{
		\includegraphics[width=3.1in]{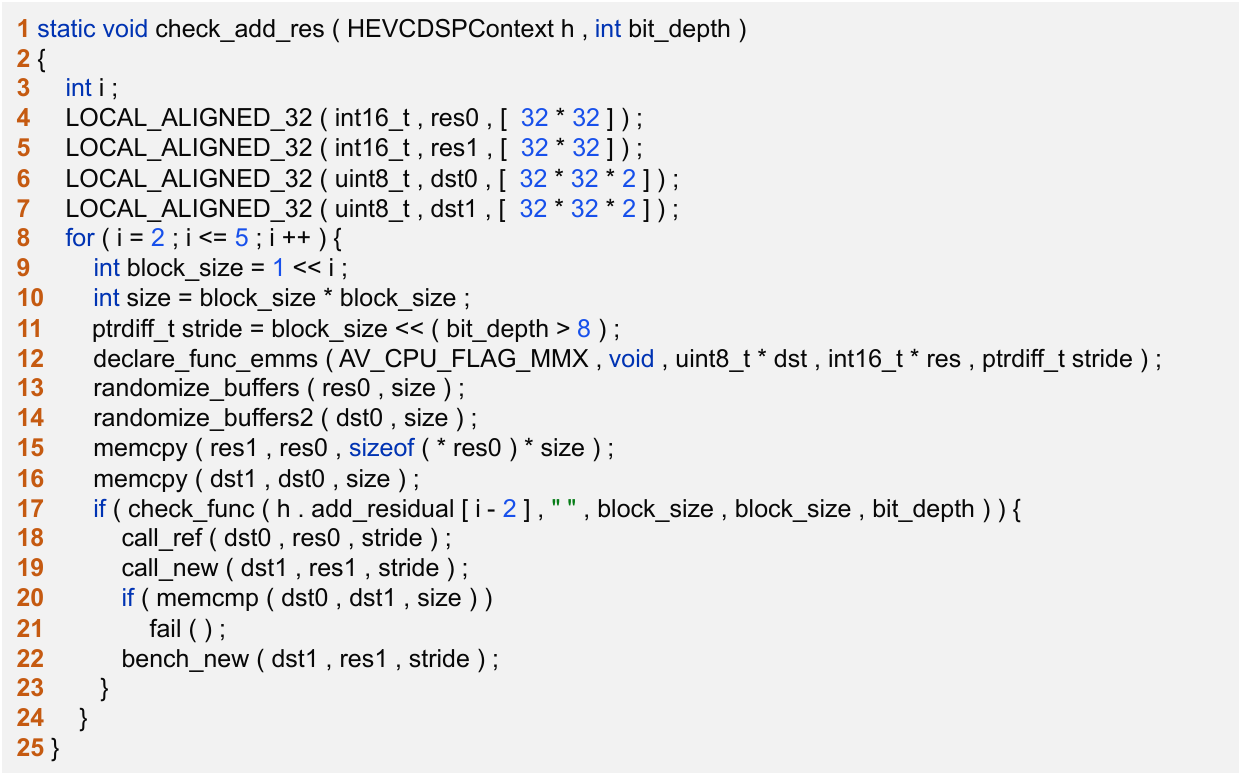}
	}
	\subfloat[Patched\label{fig:case1_fix}]
	{
		\includegraphics[width=3.1in]{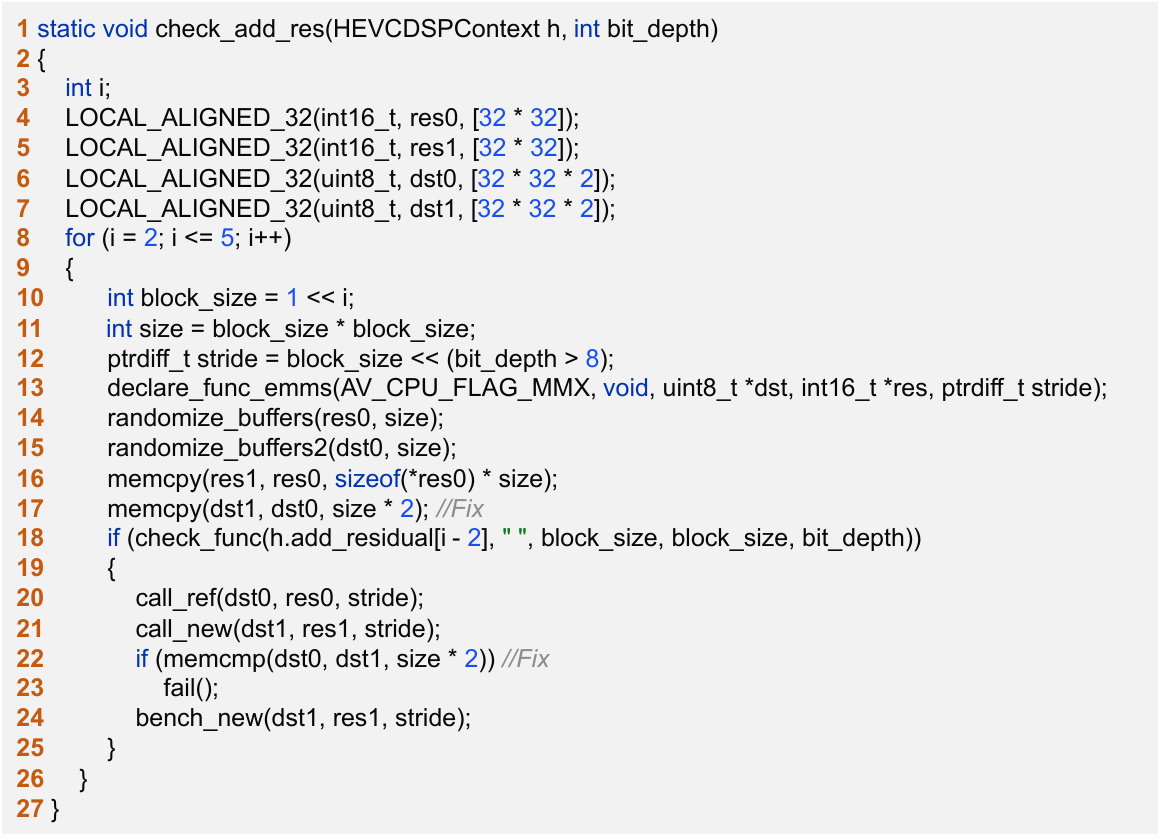}
	}
	\caption{Case 1}
	\label{fig:case1}
\end{figure*}

\begin{figure*}[!htb]
	\centering
	\subfloat[Vulnerable\label{fig:case2_vul}]{
		\includegraphics[width=3.1in]{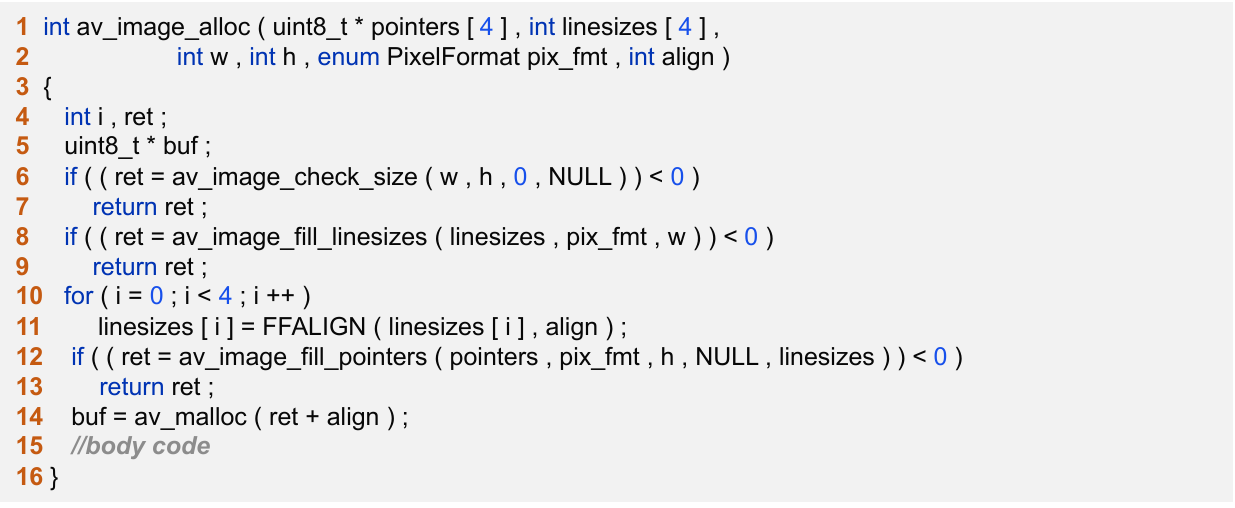}
	}
	\subfloat[Patched\label{fig:case2_fix}]
	{
		\includegraphics[width=3.1in]{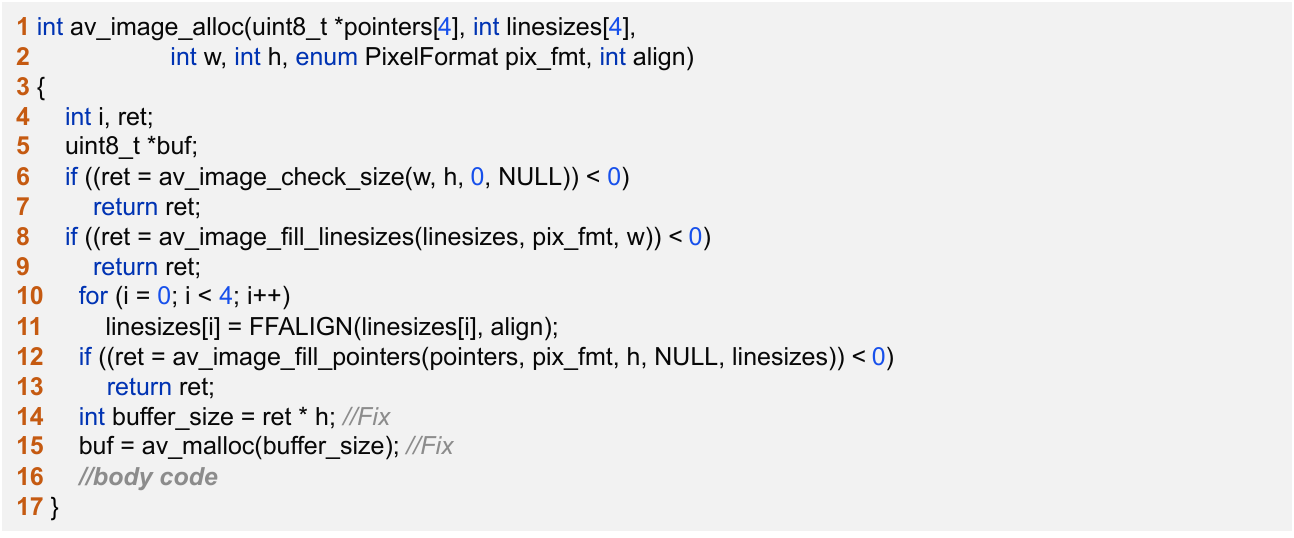}
	}
	\caption{Case 2}
	\label{fig:case2}
\end{figure*}

\begin{figure*}[!htb]
	\centering
	\subfloat[Vulnerable\label{fig:case3_vul}]{
		\includegraphics[width=3.1in]{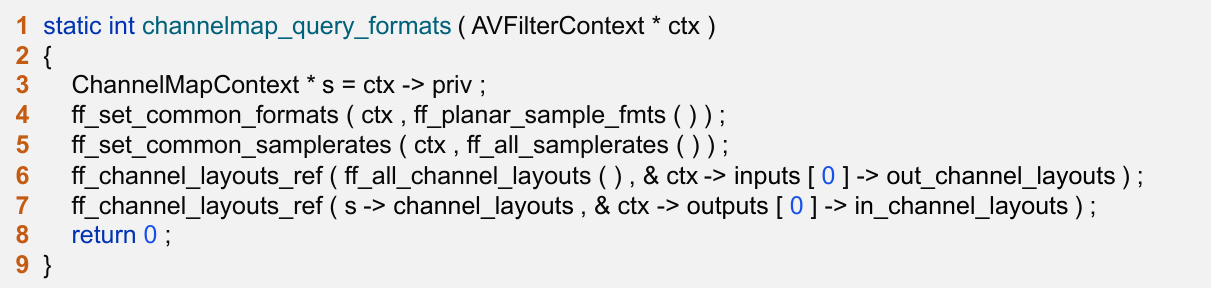}
	}
	\subfloat[Patched\label{fig:case3_fix}]
	{
		\includegraphics[width=3.1in]{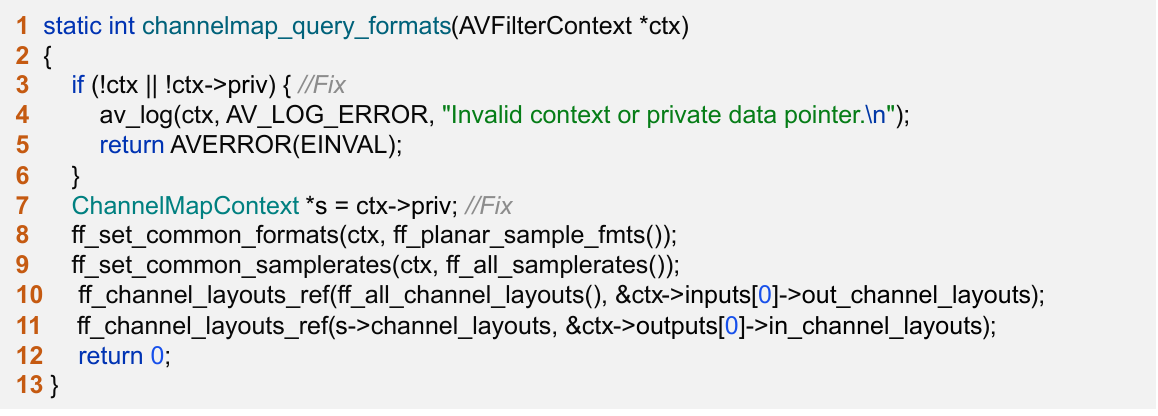}
	}
	\caption{Case 3}
	\label{fig:case3}
\end{figure*}

\begin{figure*}[!htb]
	\centering
	\subfloat[Vulnerable\label{fig:case4_vul}]{
		\includegraphics[width=3.1in]{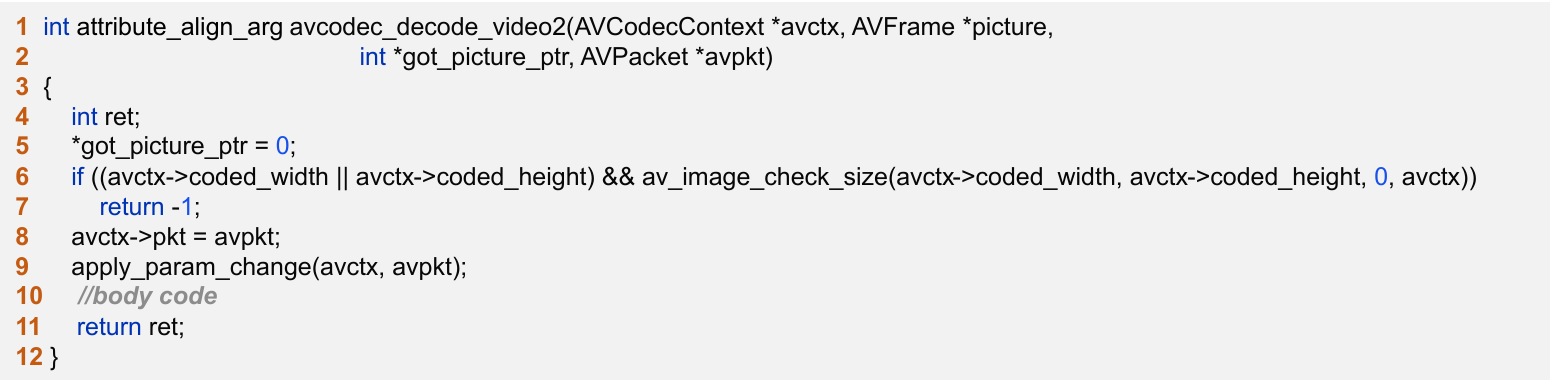}
	}
	\subfloat[Patched\label{fig:case4_fix}]
	{
		\includegraphics[width=3.1in]{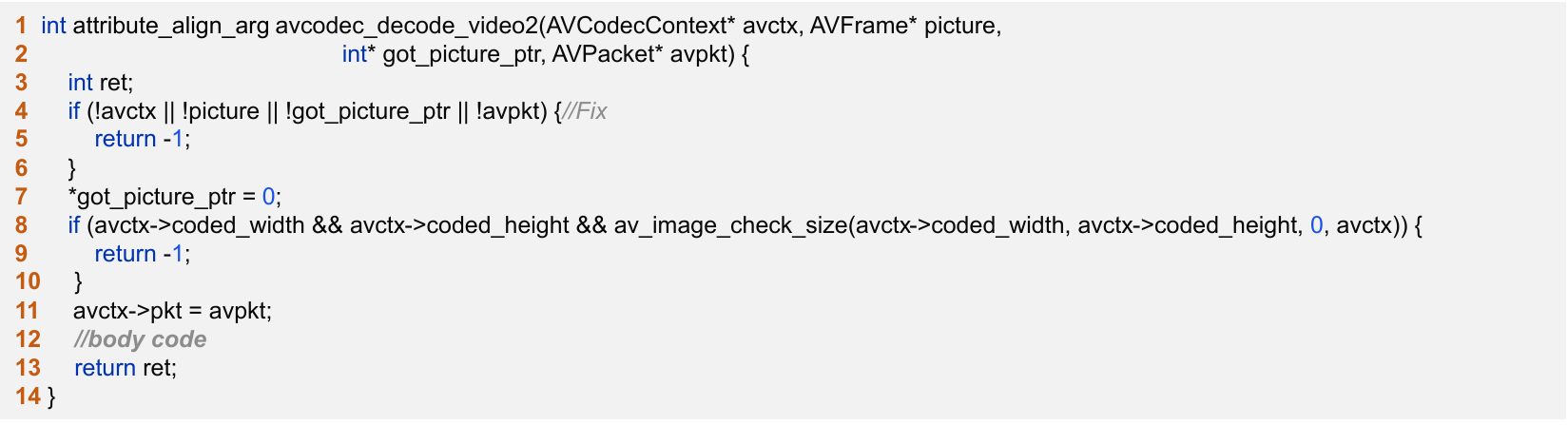}
	}
	\caption{Case 4}
	\label{fig:case4}
\end{figure*}

Case 1, as depicted in Figure \ref{fig:case1_vul}, exhibits vulnerabilities related to memory operations. Specifically, improper utilization of the \texttt{memcpy} and \texttt{memcmp} functions can lead to buffer overflow and memory comparison issues. Line 16 demonstrates the \texttt{memcpy} function copying a memory block of  \texttt{size} bytes from the source buffer \texttt{res0} to the destination buffer \texttt{res1}. However, the destination buffer is allocated for a size of \texttt{32 * 32 * 2} bytes. This mismatch in buffer sizes can lead to a buffer overflow, causing the copying operation to write data beyond the allocated memory space. This overflow has the potential to corrupt adjacent memory or even cause an application crash.

Furthermore, Line 20 introduces a vulnerability associated with the \texttt{memcmp} function. However, each of these buffers is allocated for a size of \texttt{32 * 32 * 2} bytes. This discrepancy implies that the comparison may overlook differences beyond \texttt{size} bytes, potentially failing to detect variances in the additional bytes. The corrected code in Figure \ref{fig:case1_fix} ensures proper memory operation by rectifying the allocation and copying procedures, thereby constraining them within the bounds of allocated memory buffers. Consequently, the risk of buffer overflows and memory comparison inconsistencies is considerably reduced.

The analysis of Fig.\ref{fig:case1} emphasizes the vital significance of accurate memory management in software development. Mishandling of memory functions such as \texttt{memcpy} and \texttt{memcmp} can result in severe security vulnerabilities, including buffer overflows and incorrect memory comparisons. The vulnerabilities depicted in Figure \ref{fig:case1_vul} have been effectively addressed in Figure \ref{fig:case1_fix} through appropriate fixes. DefectHunter no longer marks fixed code as a vulnerability. It is imperative for developers to maintain vigilance in applying appropriate memory-handling techniques to mitigate potential exploits and prevent security breaches.

\subsubsection{Case 2}

In Case 2 (Seeing Figure \ref{fig:case2}), the function \texttt{av\_image\_fill\_pointers} is called to populate the pointers array, returning the necessary buffer size (\texttt{ret}) for the data.

Subsequently, the code allocates memory for the buffer \texttt{buf} using \texttt{av\_malloc}, potentially resulting in an incorrect size if \texttt{align} is added to it.

The correct buffer size should be determined by multiplying \texttt{ret} (representing the required size for each line) by \texttt{h} (the number of lines).

However, as shown in Figure \ref{fig:case2_vul}, the buffer size is calculated as \texttt{ret + align}, which can be incorrect and may result in buffer overflow issues, particularly if \texttt{align} exceeds the size needed for each line.

The modification calculates the correct buffer size by multiplying \texttt{ret} (representing the size of each line) by \texttt{h} (representing the number of lines) (see Figure \ref{fig:case2_fix}). This ensures that the allocated buffer \texttt{buf} is sufficiently large to accommodate all the data without encountering any buffer overflow problems.

By incorporating this modification, the code becomes safer and less prone to potential security vulnerabilities associated with buffer overflows.

Furthermore, after implementing this fix, DefectHunter's assessment indicates that the repaired code is no longer flagged as vulnerable. DefectHunter accurately acknowledges that the corrected calculation of the buffer size reduces the likelihood of buffer overflow, thereby enhancing the overall security of the function.

\subsubsection{Case 3}

The main issue in Fig. \ref{fig:case3_vul} is the absence of proper validation checks for the pointers \texttt{ctx} and \texttt{ctx->priv}, which can potentially result in null pointer dereference or invalid memory access. This vulnerability can result in program crashes or other undesirable behavior. In Fig. \ref{fig:case3_fix}, we ensure that before utilizing \texttt{ctx} and \texttt{ctx->priv}, thorough validation checks are performed. If either pointer is found to be invalid, it generates an error along with a corresponding message. The following operations are executed only if the pointers are valid, effectively mitigating the risk of null pointer dereference or invalid memory access. After applying the fix, the code demonstrates improved robustness and safety. The introduced validation checks protect against possible runtime errors and crashes. Notably, the evaluation of DefectHunter shows a change in its perception that the corrected sections are free of any vulnerabilities.

\subsubsection{Case 4}

The vulnerability occurs in Fig.\ref{fig:case4_vul} at line 6, where \texttt{av\_image\_check\_size} is invoked without validating the pointers \texttt{avctx->coded\_width} and \texttt{avctx->coded\_height}. This omission leaves the code susceptible to null pointer dereferencing or accessing invalid memory locations. Such vulnerabilities could potentially lead to crashes, memory corruption, or unauthorized data exposure, depending on the context in which the code is executed.
To rectify the vulnerabilities, an effective remediation strategy was implemented (Seeing Fig.\ref{fig:case4_fix}). The main focus was on incorporating proper validity checks for each of the involved pointers before performing any operations. In the repaired code, prior to any processing, \texttt{avctx}, \texttt{picture}, \texttt{got\_picture\_ptr}, and \texttt{avpkt} are subjected to validation checks. Post-remediation, DefectHunter verifies that the code is free from identified vulnerabilities. It validates the proper implementation of validity checks on pointers, ensuring null pointer dereferencing or invalid memory access vulnerabilities are mitigated. DefectHunter significantly enhances code security, identifying similar vulnerabilities during the development and testing phases.

\section{Related Work}
\label{sec:related}
We categorize the related work into four distinct classifications: traditional approaches, machine learning-based approaches, deep learning-based approaches, and large language model-based approaches. Our research falls within the realm of deep learning-based methods.

\subsection{Traditional Approaches}
Traditional approaches encompass methods that are grounded in algorithmic matching and manual verification, eschewing the utilization of artificial intelligence. In the nascent stages of traditional vulnerability detection, experts manually crafted high-quality rule bases~\cite{Flawfinder, Findbugs, Checkmarx, 10.1145/3548606.3560552}. To mitigate the financial burden of constructing these rule bases, researchers introduced semi-automated techniques like taint tracking~\cite{Johns-taint-tracking, wang2019dftracker, High-Order-Taint-Style-10.1145/3460120.3484798, 10.1145/3548606.3560664, 10.1145/3548606.3559391}, symbolic execution~\cite{Baldoni-symbolic-exe, Symbolic_exe_smart_contracts}, and fuzzing~\cite{Dinh2021FavocadoFT, fuzzing_smart_contracts}, along with methods for code similarity-based vulnerability matching~\cite{VDSimilar}. These approaches achieve their objectives but often suffer from a lack of full automation and high labor costs.

\subsection{Machine Learning-Based Approaches}
Machine learning-based approaches offer automated categorization as a core feature, thereby enhancing vulnerability detection capabilities and reducing manual labor. Prominent techniques in this category include Logistic Regression, Multi-Layer Perceptron, Support Vector Machines, and Random Forest. Al-Yaseen \textit{et al.}~\cite{al2017multi} proposed a multi-level hybrid intrusion detection model employing Support Vector Machines and extreme learning machines. Additionally, they optimized the training datasets using a modified k-means algorithm. Ghaffarian \textit{et al.}~\cite{Survey-10.1145/3092566} conducted a comparative study of machine learning algorithms for vulnerability detection, highlighting the superior performance of the Random Forest algorithm. Lomio \textit{et al.}~\cite{lomio2022just} conducted an empirical study comparing the effectiveness of different machine learning approaches for vulnerability identification, and considers available metrics seem to be not enough. In addition, this paper considers that ensemble-based classifiers may perform better.  Zolanvari \textit{et al.}~\cite{8693904} assessed the suitability of machine learning-based vulnerability detection in Internet of Things (IoT) settings. While effective, these methods sometimes falter when faced with intricate challenges.

\subsection{Deep Learning-Based Approaches}
Deep learning-based methods employ intricate neural networks to augment the capability of vulnerability detection models in addressing complex issues. Li \textit{et al.} introduced Vuldeepecker~\cite{8846081}, which utilizes flat language sequences of source code to train neural networks. However, this approach sacrifices the semantic nuances of the code. Consequently, alternative research has explored the use of graph or tree representations, such as the Abstract Syntax Tree (AST) or Control Flow Graph (CFG), for training purposes. Allamanis \textit{et al.}~\cite{allamanis2017learning} suggested techniques for transforming source code into graphs and for extending gated graph neural networks. Wang \textit{et al.}~\cite{9293321} presented FUNDED, a graph-based framework for identifying vulnerabilities. Steenhoek \textit{et al.}~\cite{10172583} conducted an empirical study to examine the correlation among predictions made by different SOTP models and the relationship between dataset size and these models' performance using widely used datasets. Hin \textit{et al.}~ conducted LineVD~\cite{10.1145/3524842.3527949} based on graph neural networks, which explores the application of graph neural networks for vulnerability identification and improves the prediction performance of function codes without vulnerabilities by addressing the conflicting results between the information at the function level and the information at the statement level. Our work builds on these existing studies to enhance the performance of vulnerability detection. 

\subsection{Large Language Model-Based Approaches}
This category represents a specialized subset of deep learning-based approaches. Researchers commonly utilize pre-trained large language models, followed by domain-specific fine-tuning. Examples include Llama~\cite{Llama}, CodeX~\cite{CodeX}, ChatGPT~\cite{ChatGPT}, and GPT-4~\cite{GPT-4}. We developed the Pongo model based on the Llama architecture for vulnerability detection. Pearce \textit{et al.}~\cite{Zero-Shot-10179324} and Cheshkov \textit{et al.}~\cite{cheshkov2023evaluation} conducted studies evaluating the effectiveness of large language models in this domain. Although these models do not offer substantial advantages over general deep learning models, their extensive parameter requirements contribute to higher training costs.

\section{Conclusions and Future Research}
\label{sec:conclusion}

In this paper, we present DefectHunter, a novel model for vulnerability identification. DefectHunter extracts features from code and identifies vulnerabilities by converting the code into four distinct structural representations, which are subsequently processed through Conformer blocks. Experimental results indicate that DefectHunter sets a new technological benchmark for detecting vulnerabilities in real-world open-source projects using machine learning techniques. In addition, we conduct both ablation studies and case studies to delve further into the intricacies of the model. In the future, we envision significant potential for integrating large existing language models with our designed network and working with LangChain to establish a comprehensive knowledge base for vulnerability identification.

\section*{Acknowledgment}
This study was supported by the National Natural Science Foundation of China (62002067) and the Guangzhou Youth Talent of Science (QT20220101174).

\bibliographystyle{IEEEtran}
\bibliography{IEEE}

\begin{thebibliography}{10}
\providecommand{\url}[1]{#1}
\csname url@samestyle\endcsname
\providecommand{\newblock}{\relax}
\providecommand{\bibinfo}[2]{#2}
\providecommand{\BIBentrySTDinterwordspacing}{\spaceskip=0pt\relax}
\providecommand{\BIBentryALTinterwordstretchfactor}{4}
\providecommand{\BIBentryALTinterwordspacing}{\spaceskip=\fontdimen2\font plus
\BIBentryALTinterwordstretchfactor\fontdimen3\font minus
  \fontdimen4\font\relax}
\providecommand{\BIBforeignlanguage}[2]{{%
\expandafter\ifx\csname l@#1\endcsname\relax
\typeout{** WARNING: IEEEtran.bst: No hyphenation pattern has been}%
\typeout{** loaded for the language `#1'. Using the pattern for}%
\typeout{** the default language instead.}%
\else
\language=\csname l@#1\endcsname
\fi
#2}}
\providecommand{\BIBdecl}{\relax}
\BIBdecl

\bibitem{Flawfinder}
\BIBentryALTinterwordspacing
``Flawfinder.'' [Online]. Available: \url{https://www.dwheeler.com/flawfinder/}
\BIBentrySTDinterwordspacing

\bibitem{Findbugs}
\BIBentryALTinterwordspacing
``Findbugs.'' [Online]. Available: \url{https://findbugs.sourceforge.net/}
\BIBentrySTDinterwordspacing

\bibitem{VCCFinder-10.1145/2810103.2813604}
\BIBentryALTinterwordspacing
H.~Perl \emph{et~al.}, ``Vccfinder: Finding potential vulnerabilities in
  open-source projects to assist code audits,'' in \emph{Proceedings of the
  22nd ACM SIGSAC Conference on Computer and Communications Security}, ser. CCS
  '15.\hskip 1em plus 0.5em minus 0.4em\relax New York, NY, USA: Association
  for Computing Machinery, 2015, p. 426–437. [Online]. Available:
  \url{https://doi.org/10.1145/2810103.2813604}
\BIBentrySTDinterwordspacing

\bibitem{Survey-10.1145/3092566}
\BIBentryALTinterwordspacing
S.~M. Ghaffarian \emph{et~al.}, ``Software vulnerability analysis and discovery
  using machine-learning and data-mining techniques: A survey,'' \emph{ACM
  Comput. Surv.}, vol.~50, no.~4, aug 2017. [Online]. Available:
  \url{https://doi.org/10.1145/3092566}
\BIBentrySTDinterwordspacing

\bibitem{FUNDED}
H.~Wang \emph{et~al.}, ``Combining graph-based learning with automated data
  collection for code vulnerability detection,'' \emph{IEEE Transactions on
  Information Forensics and Security}, vol.~16, pp. 1943--1958, 2021.

\bibitem{zhou2019devign}
Y.~Zhou \emph{et~al.}, ``Devign: Effective vulnerability identification by
  learning comprehensive program semantics via graph neural networks,''
  \emph{Advances in neural information processing systems}, vol.~32, 2019.

\bibitem{CodeBert}
\BIBentryALTinterwordspacing
Z.~Feng \emph{et~al.}, ``Codebert: {A} pre-trained model for programming and
  natural languages,'' \emph{CoRR}, vol. abs/2002.08155, 2020. [Online].
  Available: \url{https://arxiv.org/abs/2002.08155}
\BIBentrySTDinterwordspacing

\bibitem{wang2021codet5}
Y.~Wang \emph{et~al.}, ``Codet5: Identifier-aware unified pre-trained
  encoder-decoder models for code understanding and generation,'' in
  \emph{Proceedings of the 2021 Conference on Empirical Methods in Natural
  Language Processing}, 2021, pp. 8696--8708.

\bibitem{GPT-4}
\BIBentryALTinterwordspacing
``Gpt-4.'' [Online]. Available: \url{https://openai.com/gpt-4}
\BIBentrySTDinterwordspacing

\bibitem{gulati2020conformer}
A.~Gulati \emph{et~al.}, ``Conformer: Convolution-augmented transformer for
  speech recognition,'' 2020.

\bibitem{Atten6447854:online}
E.~Miller, ``Attention is off by one evan miller,''
  \url{https://www.evanmiller.org/attention-is-off-by-one.html}, 07 2023,
  (undefined 5/8/2023 13:48).

\bibitem{DBLP:journals/corr/VaswaniSPUJGKP17}
\BIBentryALTinterwordspacing
A.~Vaswani \emph{et~al.}, ``Attention is all you need,'' \emph{CoRR}, vol.
  abs/1706.03762, 2017. [Online]. Available:
  \url{http://arxiv.org/abs/1706.03762}
\BIBentrySTDinterwordspacing

\bibitem{wolf-etal-2020-transformers}
\BIBentryALTinterwordspacing
T.~Wolf \emph{et~al.}, ``Transformers: State-of-the-art natural language
  processing,'' in \emph{Proceedings of the 2020 Conference on Empirical
  Methods in Natural Language Processing: System Demonstrations}.\hskip 1em
  plus 0.5em minus 0.4em\relax Online: Association for Computational
  Linguistics, Oct. 2020, pp. 38--45. [Online]. Available:
  \url{https://www.aclweb.org/anthology/2020.emnlp-demos.6}
\BIBentrySTDinterwordspacing

\bibitem{guo2022unixcoder}
D.~Guo \emph{et~al.}, ``Unixcoder: Unified cross-modal pre-training for code
  representation,'' 2022.

\bibitem{bird2009natural}
S.~Bird \emph{et~al.}, \emph{Natural language processing with Python: analyzing
  text with the natural language toolkit}.\hskip 1em plus 0.5em minus
  0.4em\relax " O'Reilly Media, Inc.", 2009.

\bibitem{chollet2015keras}
F.~Chollet \emph{et~al.}, ``Keras,'' \url{https://keras.io}, 2015.

\bibitem{NISTSoft48:online}
``Nist software assurance reference dataset,''
  \url{https://samate.nist.gov/SARD/}, (Accessed on 09/04/2022).

\bibitem{kim-2014-convolutional}
\BIBentryALTinterwordspacing
Y.~Kim, ``Convolutional neural networks for sentence classification,'' in
  \emph{Proceedings of the 2014 Conference on Empirical Methods in Natural
  Language Processing ({EMNLP})}.\hskip 1em plus 0.5em minus 0.4em\relax Doha,
  Qatar: Association for Computational Linguistics, oct 2014, pp. 1746--1751.
  [Online]. Available: \url{https://aclanthology.org/D14-1181}
\BIBentrySTDinterwordspacing

\bibitem{li2018vuldeepecker}
\BIBentryALTinterwordspacing
Z.~Li \emph{et~al.}, ``Vuldeepecker: {A} deep learning-based system for
  vulnerability detection,'' in \emph{25th Annual Network and Distributed
  System Security Symposium, {NDSS} 2018, San Diego, California, USA, February
  18-21, 2018}.\hskip 1em plus 0.5em minus 0.4em\relax The Internet Society,
  2018. [Online]. Available:
  \url{http://wp.internetsociety.org/ndss/wp-content/uploads/sites/25/2018/02/ndss2018\_03A-2\_Li\_paper.pdf}
\BIBentrySTDinterwordspacing

\bibitem{WANG202315}
\BIBentryALTinterwordspacing
J.~Wang \emph{et~al.}, ``Deepvulseeker: A novel vulnerability identification
  framework via code graph structure and pre-training mechanism,'' \emph{Future
  Generation Computer Systems}, vol. 148, pp. 15--26, 2023. [Online].
  Available:
  \url{https://www.sciencedirect.com/science/article/pii/S0167739X23001978}
\BIBentrySTDinterwordspacing

\bibitem{chen2021evaluating}
M.~Chen \emph{et~al.}, ``Evaluating large language models trained on code,''
  2021.

\bibitem{touvron2023llama}
H.~Touvron \emph{et~al.}, ``Llama 2: Open foundation and fine-tuned chat
  models,'' 2023.

\bibitem{dettmers2023qlora}
T.~Dettmers \emph{et~al.}, ``Qlora: Efficient finetuning of quantized llms,''
  2023.

\bibitem{Checkmarx}
\BIBentryALTinterwordspacing
``Checkmarx.'' [Online]. Available: \url{https://www.checkmarx.com/}
\BIBentrySTDinterwordspacing

\bibitem{10.1145/3548606.3560552}
\BIBentryALTinterwordspacing
S.~Cui \emph{et~al.}, ``Vrust: Automated vulnerability detection for solana
  smart contracts,'' in \emph{Proceedings of the 2022 ACM SIGSAC Conference on
  Computer and Communications Security}, ser. CCS '22.\hskip 1em plus 0.5em
  minus 0.4em\relax New York, NY, USA: Association for Computing Machinery,
  2022, p. 639–652. [Online]. Available:
  \url{https://doi.org/10.1145/3548606.3560552}
\BIBentrySTDinterwordspacing

\bibitem{Johns-taint-tracking}
\BIBentryALTinterwordspacing
M.~Johns \emph{et~al.}, ``End-to-end taint tracking for detection and
  mitigation of injection vulnerabilities in web applications,'' \emph{US
  Patent 10,129,285}, 2018, query date: 2023-09-15 10:23:18. [Online].
  Available: \url{https://patents.google.com/patent/US10129285B2/en}
\BIBentrySTDinterwordspacing

\bibitem{wang2019dftracker}
P.~Wang \emph{et~al.}, ``Dftracker: detecting double-fetch bugs by multi-taint
  parallel tracking,'' \emph{Frontiers of Computer Science}, vol.~13, pp.
  247--263, 2019.

\bibitem{High-Order-Taint-Style-10.1145/3460120.3484798}
\BIBentryALTinterwordspacing
H.~Zhang \emph{et~al.}, ``Statically discovering high-order taint style
  vulnerabilities in os kernels,'' in \emph{Proceedings of the 2021 ACM SIGSAC
  Conference on Computer and Communications Security}, ser. CCS '21.\hskip 1em
  plus 0.5em minus 0.4em\relax New York, NY, USA: Association for Computing
  Machinery, 2021, p. 811–824. [Online]. Available:
  \url{https://doi.org/10.1145/3460120.3484798}
\BIBentrySTDinterwordspacing

\bibitem{10.1145/3548606.3560664}
\BIBentryALTinterwordspacing
W.~Kang \emph{et~al.}, ``Tracer: Signature-based static analysis for detecting
  recurring vulnerabilities,'' in \emph{Proceedings of the 2022 ACM SIGSAC
  Conference on Computer and Communications Security}, ser. CCS '22.\hskip 1em
  plus 0.5em minus 0.4em\relax New York, NY, USA: Association for Computing
  Machinery, 2022, p. 1695–1708. [Online]. Available:
  \url{https://doi.org/10.1145/3548606.3560664}
\BIBentrySTDinterwordspacing

\bibitem{10.1145/3548606.3559391}
\BIBentryALTinterwordspacing
C.~Luo \emph{et~al.}, ``Tchecker: Precise static inter-procedural analysis for
  detecting taint-style vulnerabilities in php applications,'' in
  \emph{Proceedings of the 2022 ACM SIGSAC Conference on Computer and
  Communications Security}, ser. CCS '22.\hskip 1em plus 0.5em minus
  0.4em\relax New York, NY, USA: Association for Computing Machinery, 2022, p.
  2175–2188. [Online]. Available:
  \url{https://doi.org/10.1145/3548606.3559391}
\BIBentrySTDinterwordspacing

\bibitem{Baldoni-symbolic-exe}
\BIBentryALTinterwordspacing
R.~Baldoni \emph{et~al.}, ``A survey of symbolic execution techniques,''
  \emph{ACM Computing Surveys (CSUR)}, 2018, query date: 2023-09-15 10:28:28.
  [Online]. Available: \url{https://dl.acm.org/doi/abs/10.1145/3182657}
\BIBentrySTDinterwordspacing

\bibitem{Symbolic_exe_smart_contracts}
\BIBentryALTinterwordspacing
D.~Wang \emph{et~al.}, ``Wana: Symbolic execution of wasm bytecode for
  cross-platform smart contract vulnerability detection,'' \emph{arXiv preprint
  arXiv:2007.15510}, 2020, query date: 2023-09-15 10:28:28. [Online].
  Available: \url{https://arxiv.org/abs/2007.15510}
\BIBentrySTDinterwordspacing

\bibitem{Dinh2021FavocadoFT}
\BIBentryALTinterwordspacing
S.~T. Dinh \emph{et~al.}, ``Favocado: Fuzzing the binding code of javascript
  engines using semantically correct test cases,'' in \emph{Network and
  Distributed System Security Symposium}, 2021. [Online]. Available:
  \url{https://api.semanticscholar.org/CorpusID:231591466}
\BIBentrySTDinterwordspacing

\bibitem{fuzzing_smart_contracts}
J.~He \emph{et~al.}, ``Learning to fuzz from symbolic execution with
  application to smart contracts,'' in \emph{Proceedings of the 2019 ACM SIGSAC
  Conference on Computer and Communications Security}, 2019, pp. 531--548.

\bibitem{VDSimilar}
\BIBentryALTinterwordspacing
H.~Sun \emph{et~al.}, ``Vdsimilar: Vulnerability detection based on code
  similarity of vulnerabilities and patches,'' \emph{Computers \&Security},
  2021, query date: 2023-09-15 10:40:40. [Online]. Available:
  \url{https://www.sciencedirect.com/science/article/pii/S0167404821002418}
\BIBentrySTDinterwordspacing

\bibitem{al2017multi}
W.~L. Al-Yaseen \emph{et~al.}, ``Multi-level hybrid support vector machine and
  extreme learning machine based on modified k-means for intrusion detection
  system,'' \emph{Expert Systems with Applications}, vol.~67, pp. 296--303,
  2017.

\bibitem{lomio2022just}
F.~Lomio \emph{et~al.}, ``Just-in-time software vulnerability detection: Are we
  there yet?'' \emph{Journal of Systems and Software}, vol. 188, p. 111283,
  2022.

\bibitem{8693904}
M.~Zolanvari \emph{et~al.}, ``Machine learning-based network vulnerability
  analysis of industrial internet of things,'' \emph{IEEE Internet of Things
  Journal}, vol.~6, no.~4, pp. 6822--6834, Aug 2019.

\bibitem{8846081}
D.~Zou \emph{et~al.}, ``Vuldeepecker: A deep learning-based system for
  multiclass vulnerability detection,'' \emph{IEEE Transactions on Dependable
  and Secure Computing}, vol.~18, no.~5, pp. 2224--2236, 2021.

\bibitem{allamanis2017learning}
M.~Allamanis \emph{et~al.}, ``Learning to represent programs with graphs,''
  \emph{arXiv preprint arXiv:1711.00740}, 2017.

\bibitem{9293321}
H.~Wang \emph{et~al.}, ``Combining graph-based learning with automated data
  collection for code vulnerability detection,'' \emph{IEEE Transactions on
  Information Forensics and Security}, vol.~16, pp. 1943--1958, 2021.

\bibitem{10172583}
B.~Steenhoek \emph{et~al.}, ``An empirical study of deep learning models for
  vulnerability detection,'' in \emph{2023 IEEE/ACM 45th International
  Conference on Software Engineering (ICSE)}, 2023, pp. 2237--2248.

\bibitem{10.1145/3524842.3527949}
\BIBentryALTinterwordspacing
D.~Hin \emph{et~al.}, ``Linevd: Statement-level vulnerability detection using
  graph neural networks,'' in \emph{Proceedings of the 19th International
  Conference on Mining Software Repositories}, ser. MSR '22.\hskip 1em plus
  0.5em minus 0.4em\relax New York, NY, USA: Association for Computing
  Machinery, 2022, p. 596–607. [Online]. Available:
  \url{https://doi.org/10.1145/3524842.3527949}
\BIBentrySTDinterwordspacing

\bibitem{Llama}
\BIBentryALTinterwordspacing
``Llama.'' [Online]. Available: \url{https://ai.meta.com/llama/}
\BIBentrySTDinterwordspacing

\bibitem{CodeX}
\BIBentryALTinterwordspacing
``Codex.'' [Online]. Available: \url{https://openai.com/blog/openai-codex/}
\BIBentrySTDinterwordspacing

\bibitem{ChatGPT}
\BIBentryALTinterwordspacing
``Chatgpt.'' [Online]. Available: \url{https://chat.openai.com}
\BIBentrySTDinterwordspacing

\bibitem{Zero-Shot-10179324}
H.~Pearce \emph{et~al.}, ``Examining zero-shot vulnerability repair with large
  language models,'' in \emph{2023 IEEE Symposium on Security and Privacy
  (SP)}, 2023, pp. 2339--2356.

\bibitem{cheshkov2023evaluation}
A.~Cheshkov \emph{et~al.}, ``Evaluation of chatgpt model for vulnerability
  detection,'' \emph{arXiv preprint arXiv:2304.07232}, 2023.

\end{thebibliography}

\end{document}